\documentclass[twocolumn,secnumarabic,amssymb, nobibnotes, aps]{revtex4-1}
\usepackage{graphicx} 
\usepackage{booktabs}
\usepackage{natbib}
\usepackage{color}
\usepackage[colorlinks=true, allcolors=blue]{hyperref}
\usepackage{hyperref}

\begin{document}
\title{Tunable Half-Metallicity and Edge Magnetism of H-saturated InSe Nanoribbons}%
\author{Weiqing Zhou}%
\author{Guodong Yu}%
\author{A. N. Rudenko}
\affiliation{School of Physics and Technology, Wuhan University, Wuhan, 430072, People’s Republic of China}
\author{Shengjun Yuan}%
\email{s.yuan@whu.edu.cn}
\affiliation{School of Physics and Technology, Wuhan University, Wuhan, 430072, People’s Republic of China}
\date{\today}
\begin{abstract}
	We report on a theoretical study of electronic and magnetic properties of hydrogen-saturated InSe nanoribbons~(H-ZISNs). Based on hybrid-functional first-principles calculations, we find that H-ZISNs exhibit tunable half-metallicity and short-range ferromagnetic order. We first show that $p$-type doping turns narrow H-ZISNs from semimetal to half-metal with spin-polarization along In-terminated edge. This behavior is further analyzed in terms of a two-band tight-binding model, which provides a tractable description of the H-ZISN electronic structure, and serves as a starting point for the determination of magnetic interactions. The dominant exchange interaction determined within the Heisenberg model is found to be ferromagnetic independently of the ribbon width and charge doping. Short-range stability of magnetic order is assessed in terms of the zero-field spin correlation length, which is found to be about 1 nm at liquid nitrogen temperatures. Finally, by calculating spin-dependent transport properties, we find a doping regime in which strongly spin-selective electrical conductivities can be observed. Our findings suggest that H-ZISNs are appealing candidates for the realization of spintronic effects at the nanoscale.    
\end{abstract}
\maketitle

\section{Introduction}
 The possibility of magnetic ordering in two dimensional (2D) materials is among the most topical problems in physics since the discovery of graphene. Edge termination, one of the most promising ways to induce magnetism in 2D systems, has long been a subject of interest \cite{fujita1996peculiar,kan2008half,li2009spin,2014roomtemperature,2006energygap,son2006half,wang2009ndoping,xu2010edgesaturated}. For instance, zigzag graphene nanoribbons (ZGNRs) with narrow widths are predicted to be antiferromagnetic semiconductors \cite{fujita1996peculiar,2006energygap}, while for larger widths ($>$7~nm), a semiconductor-to-metal transition is observed experimentally \cite{2014roomtemperature}. A number of methods can be applied to control magnetism of ZGNRs, among which are doping and edge-modification \cite{kan2008half,li2009spin,xu2010edgesaturated}. Edge magnetism has also been proposed in other 2D materials such as MoS$_2$ \cite{MoS2nanoribbons}, black phosphorus (BP) \cite{BPnanoribbons} and ZnO \cite{ZnOnanoribbons}. One of the most interesting aspects of edge magnetism is half-metallicity. In half-metals, spin polarization results in the coexistence of metallic nature for electrons with one spin orientation and insulating nature for electrons with the other. It has been proposed that half-metallicity in ZNGRs can be realized under external transverse electric field \cite{son2006half}. However, for graphene, ZNGRs are less stable compared to some other nonmagnetic structures, such as mono- and dihydrogenated armchair nanoribbons \cite{kunstmann2011stability,wassmann2008structure}, as indicated by spontaneous reconstructions of zigzag edges at room temperature \cite{koskinen2008self}. For BP, the environmental instability limits its application in nanoscale electronic and magnetic devices \cite{island2015environmentalBP}. To find magnetic materials with structural and environmental stability remains an important topic in 2D spintronics.  

Monolayer InSe, a new member in the family of 2D materials, has been fabricated very recently \cite{bandurin2017InSe}. It is an indirect gap semiconductor with an optical gap of around 2.9 eV \cite{bandurin2017InSe}. The carrier mobility of multilayer InSe exceeds $10^{3}$ and $10^{4}$~cm$^{2}$V$^{-1}$s$^{-1}$ at room and liquid helium temperatures, respectively \cite{bandurin2017InSe}. Interestingly, the effective masses in InSe are weekly dependent on the layer thickness and are significantly smaller than those in other van der Waals crystals \cite{mudd2016direct}. InSe exhibits a higher environmental stability than few-layer BP, and a higher room-temperature mobility than few-layer transitional metal dichalcogenides \cite{bandurin2017InSe}. In addition, few-layer InSe has an extremely strong photoresponse and fast response time \cite{bandurin2017InSe,mudd2016direct,lei2014evolution}, which make it a promising candidate for optoelectronic applications. Besides, half-metallicity in 2D halogen atom adsorbed InSe-X (X=F, Cl, Br and I) nanosheet has been predicted from first-principles calculations \cite{liu2018graphene}.

Very recently, density functional theory (DFT) studies point to the possibility of edge magnetic ordering in InSe nanoribbon \cite{wu2018modulation}. Similar to graphene, edge spin-polarization is predicted for all zigzag nanoribbons with/without hydrogen or halogen saturation. H-saturated zigzag InSe nanoribbons (H-ZISNs) are semimetals with electrons and holes existing at Se- and In-terminated edges, respectively. Interestingly, the spin-polarization in H-ZISNs is only localized along the In-terminated edge. Within DFT, magnetic ground state of H-ZISNs is found to be ferromagnetic (FM) due to its lower total energy compared to the nonmagnetic configuration. Single-edge magnetic polarization appears appealing in the context of obtaining single-edge spin current by means of the cutting edge technologies. 
Interestingly, magnetism in InSe nanoribbons is predicted to be robust with respect to the passivation by hydrogen, which is important from the point of view of practical applications. Neither origin nor stability of edge magnetism in InSe has been yet analyzed in detail. Practical potential of this phenomenon also remains unclear.

In this paper, we discuss magnetic and electronic properties of H-terminated zigzag InSe nanoribbons (H-ZISNs). We focus on tunable half-metallicity and intrinsic magnetism at the nanometer scale, and their potential applications. We start from first-principles electronic structure calculations at the hybrid-functional level, and construct a tight-binding (TB) Hamiltonian. In order to analyze magnetic properties in more details, we estimate the exchange interaction parameters in H-ZISNs defined in terms of the isotropic Heisenberg model. We then determine the magnetic ground state and spin correlation length as function of the nanoribbon width and temperature. Finally, we calculate spin-dependent electron conductivities as a function of doping, and discuss the potential application of H-ZISNs in spintronic devices.

The paper is organized as follows. In Sec.~II, we present first-principles electronic structure of $N$-H-ZISNs ($N$=5--12) calculated using DFT-HSE06 and discuss their half-metallicity. A simplified description of the electronic structure within a two-band TB model is given in Sec.~III. In Sec.~IV, we estimate exchange interactions, and parametrize the Heisenberg model to determine the ground magnetic state of $N$-H-ZISNs. Having obtained the Heisenberg model, in Sec.~V, we calculate the spin correlation length dependence on the nanoribbon width and temperature, as well as discuss fundamental limits of H-ZISNs magnetic applications. In Sec.~VI, we discuss the spin-dependent conductivity as a function of chemical potential (doping) in H-ZISNs. In Sec.~VII, we summarize our findings and conclude the paper.

\section{First-principles calculations}
\begin{figure}[ht]	
	\includegraphics[width=8.5cm]{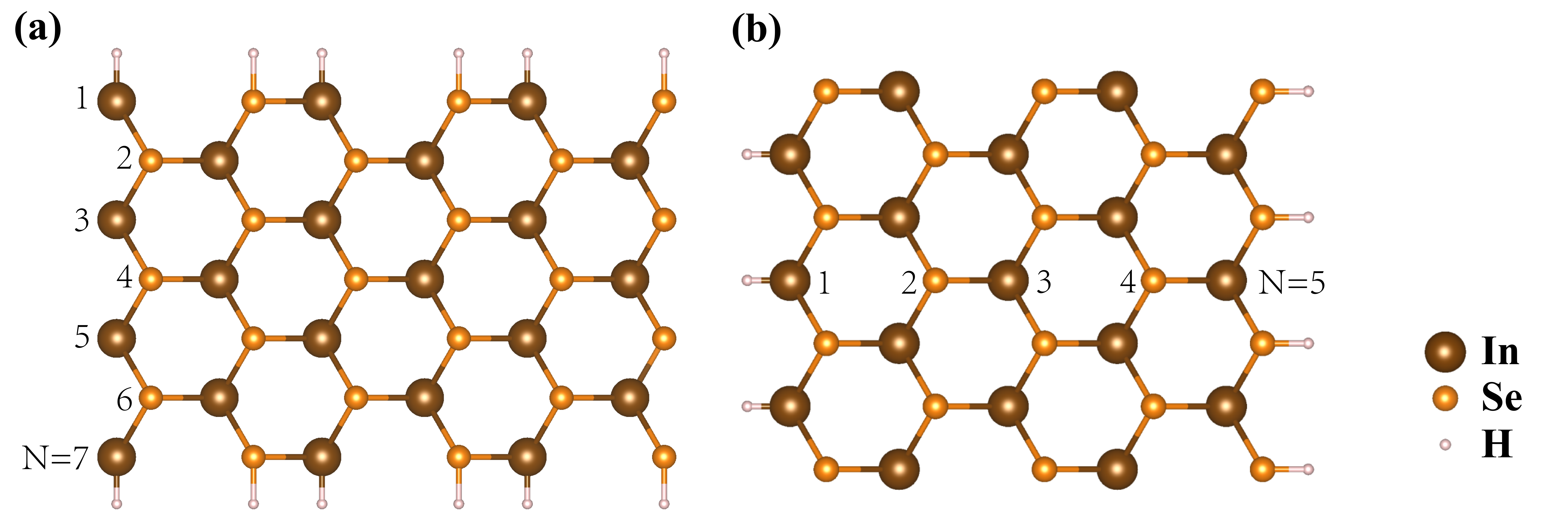}
	\caption{Schematic crystal structures of (a) 7-H-AISN and (b) 5-H-ZISN.}
	\label{fig:label1}
\end{figure}

Monolayer InSe has a hexagonal structure with $D_{3h}$ point group, which includes a 3-fold rotation symmetry axis ($C_{3v}$), and a mirror plane ($\sigma_h$). Two types of nanoribbons can be constructed from InSe, namely, nanoribbons with zigzag (ZISN) and armchair (AISN) edges. To avoid the presence of highly reactive dangling bonds, we saturate edge atoms by hydrogen (see Fig.~\ref{fig:label1}). We label the corresponding nanoribbons as H-ZISNs and H-AISNs, respectively. In contrast to H-AISNs, where vertical mirror plane ($\sigma_v$) is present, opposite edges of H-ZISNs are terminated by different (In and Se) atoms, which results in the structure with lower symmetry.
\begin{figure}[ht]	
	\includegraphics[width=8.5cm]{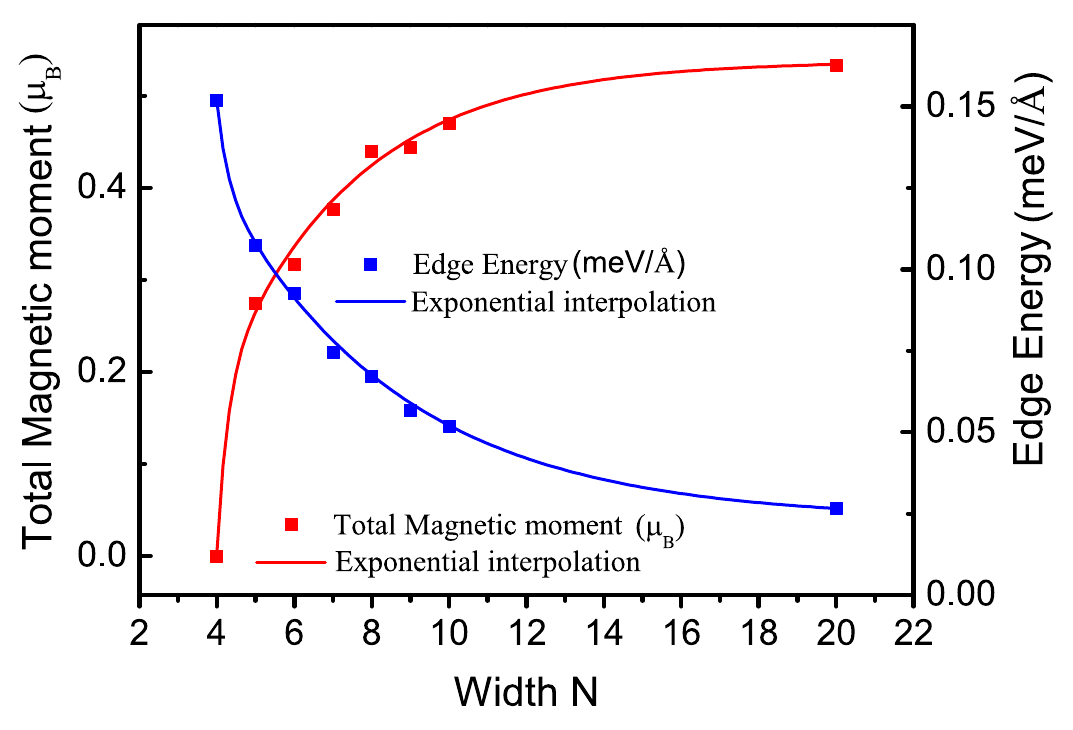}
	\caption{Edge energy and total magnetic moment of H-ZISNs calculated for different widths $N$. Both quantities converge as $N$ increases. The total magnetic moment and edge energy are expected to be $0.54~\mu_{B}$ and $23.2$~meV in the large-$N$ limit.}
	\label{fig:label2}
\end{figure}

Compared to H-ZISNs, all armchair InSe nanoribbons are predicted to be nonmagnetic semiconductors with a direct band gap of $\sim$1~eV at the $\Gamma$ point \citep{wu2018modulation}. Here, we mainly focus on magnetic properties and, therefore, consider only H-ZISNs. Electronic and magnetic properties are studied in our paper by performing DFT calculations using the \emph{Vienna Ab inito simulation Package} ({\sc VASP}), which implements the projected augmented wave method \cite{blochl1994PAW,kresse1996plane,kresse1999ultrasoft,perdew1996generalized}. 
$4d^{10}5s^{2}5d^{1}$ electrons of In and $4s^{2}4p^{4}$ of Se were treated as valence electrons. The energy cutoff of 500~eV was set for a plane-wave basis set. The reciprocal space was sampled by $21$ k-points for all InSe nanoribbons. The vacuum space of at least $10$ {\AA} was introduced in the directions perpendicular to the ribbons to avoid spurious interactions between periodic supercell images. All structures were relaxed until the residual force on each atom were less than $0.01$~eV/\AA. Magnetic moment of each atom was calculated by the Bader electron population analysis method \cite{henkelman2006baber}. 

As a starting point, we use the generalized gradient approximation (GGA) as proposed by Perdew, Burke, and Ernzerhof (PBE) \cite{perdew1996generalized} to describe exchange-correlation effects. 
However, it is well known that PBE functionals systematically underestimate band gaps in insulators and semiconductors \cite{tran2009accurate,heyd2004efficient}. To get a more accurate electronic structure, we adopt the hybrid functional method of Heyd, Scuseria, and Ernzerhof (HSE06) \cite{heyd2004efficient,heyd2003hybrid}. In this method, $25$\% of the exact screened Hartree-Fock (HF) exchange is incorporated into PBE exchange \cite{krukau2006influence}. 
H-ZISNs with widths $N$ 4--20 and 5--14 were calculated using the PBE and HSE06 functionals, respectively.

\begin{figure}[ht]	
	\includegraphics[width=9cm]{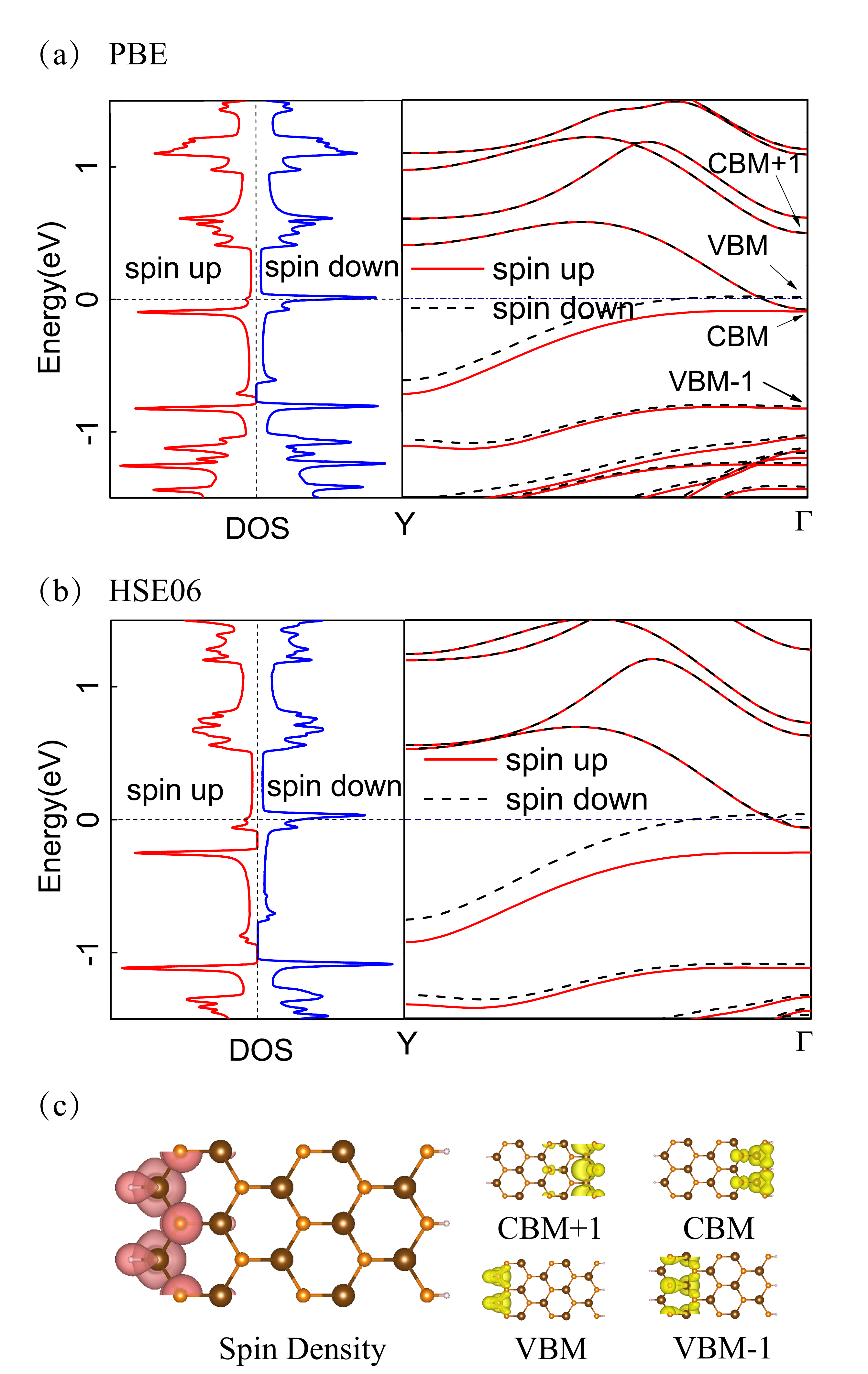}
	\caption{DOS and band structure of $5$-H-ZISN calculated using (a) PBE and (b) HSE06 methods. (c) Real-space distribution of spin density $\rho^s({\bf r})$ averaged over valence states of 5-H-ZISN, and charge density of the four relevant spin-up states at the $\Gamma$ point, $\rho_n^{\uparrow}({\bf r})$.  DOS, $g(E)=\sum_{n,k}{\delta(E-E_{n,k})}$ is calculated on a grid of interpolated 10$^4$ k-points, adopting simple Gaussian smearing with the variance $\sigma=1$ meV.}
	\label{fig:label3}
\end{figure}
\begin{figure*}[pht]		
	\includegraphics[width=18cm]{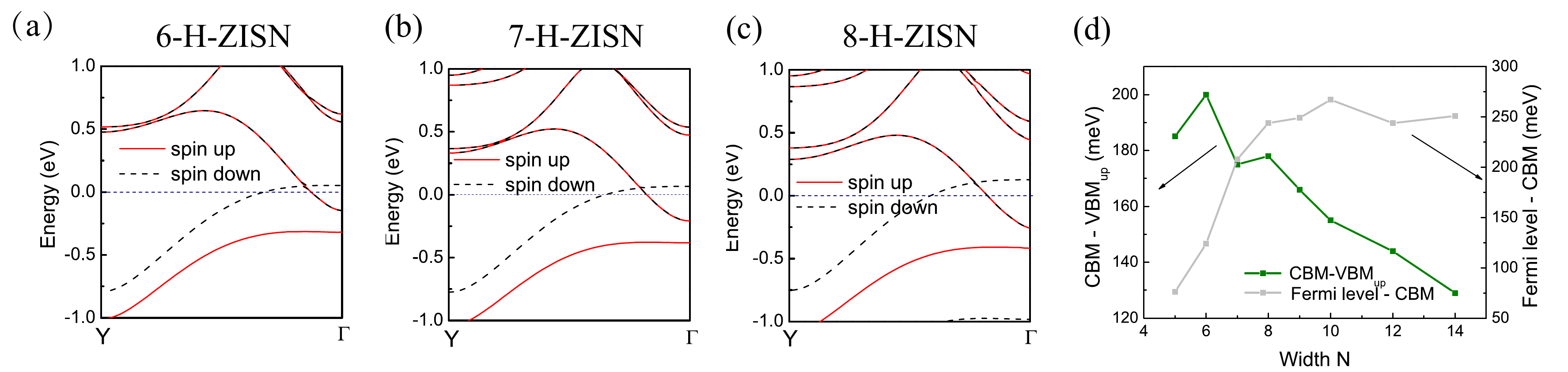}
	\caption{HSE06 band structures of (a) $6$-H-ZISN, (b) $7$-H-ZISN, and (c) $8$-H-ZISN. (d) The energy difference between CBM and spin up VBM (green line), and the energy difference between the Fermi energy and CBM (grey line).}
	\label{fig:label4}
\end{figure*}

Width-dependent magnetization and edge energy have been calculated at the PBE level to find the width $N$ at which electronic and magnetic properties converge with respect to the ribbon size. We first check the stability of two types of edges by calculating the edge energy density \cite{northrup1989energetics}, defined as $E_{edge}=(E_{tot}-\sum_{i}n_{i}\mu_{i})/2L$, where $E_{tot}$ is the total energy of the ribbon, $n_{i}$ and $\mu_{i}$ are the number and chemical potential of the $i$-th atom in the ribbon (including In, Se and H atoms), and $L$ is the unit cell length in the periodic direction. Here, H-ZISNs are treated as a reaction product of monolayer InSe and hydrogen gas. The chemical potential of In and Se atoms are taken from monolayer InSe, while the one of H atom is the total energy per atom of H$_{2}$ molecule. Because In- and Se-terminated zigzag edges always coexist, the calculated edge energy density in H-ZISNs is arithmetically averaged over two different zigzag edges. By varying the ribbon width ($N$), we find that the edge energies of $N$-H-ZISNs decrease monotonically as $N$ increases, and converge to $23.2$~meV/\AA{ } in the limit of large $N$ (see Fig.~\ref{fig:label2}). A rather small edge energy, which is comparable to graphene and MoS$_{2}$, indicates that InSe nanoribbons can be fabricated in the laboratory \cite{MoS2nanoribbons,lu2010excess,huang2009quantum}. The width-dependent magnetic moment of H-ZISNs is shown in Fig.~\ref{fig:label2}. One can see that magnetization is essentially converged at $N>10$. 

To analyze electronic properties of H-ZISNs, we first consider the smallest magnetic nanoribbon, which is 5-H-ZISN. 
As it is shown in Figs.~\ref{fig:label3} (a) and (b), 5-H-ZISN is a semimetal with finite density of states (DOS) at the Fermi energy. 
While the conduction band remains doubly degenerate, the valence band split into two bands, corresponding to different spin projections (labeled in Fig.~\ref{fig:label3} as spin up and spin down). As a results of the spin splitting, there is a direct energy gap emerging at the $\Gamma$ point for the spin up channel, whereas the spin down channel remains gapless in the vicinity of the Fermi energy. 
The value of this energy gap is about $20$~meV calculated within PBE, while it increases to $200$~meV at the HSE06 level, which is due to the larger spin splitting. The fact that hybrid functional results in the larger spin splitting indicates an important role of the exact Hartree-Fock exchange in the formation of magnetic properties of H-ZISNs, which is common in $sp$-magnetism \cite{mazurenko2016role}. The Fermi energy in pristine $5$-ZISN calculated within HSE06 is found to be about $70$~meV above the band gap, which allows us to expect the possibility to turn $5$-ZISN from semi-metal to half-metal by varying the carrier concentration via either the electric field effect or by $p$-type doping. 
Fig.~\ref{fig:label3} (c) shows real space distribution of the spin density $\rho^s({\bf r})=\sum_{n,k}(|\phi^{\uparrow}_{nk}({\bf r})|^2-|\phi^{\downarrow}_{nk}({\bf r})|^2)$ averaged over the valence bands, and the charge density of the spin-up states at the $\Gamma$ point $\rho^{\uparrow}_n({\bf r})=|\phi^{\uparrow}_{n\Gamma}({\bf r})|^2$, corresponding to two highest valence bands and two lowest conduction bands. One can see that the conduction and valence states are predominantly localized along Se- and In-terminated edges, respectively. Since only the valence band exhibits spin splitting, spin density is expectedly localized along In-terminated edge.

In order to verify generality of our findings, we perform HSE06 calculations of H-ZISNs for different ribbon widths. From our calculations it follows that a gap above $100$~meV is found between the conduction band minimum (CBM) and spin down valence band maximum (VBM) in narrow H-ZISNs when $N$ is ranging from $5$ to $14$. The trend can be seen from Fig.~\ref{fig:label4}, where the electronic bands of $6$-H-ZISN, $7$-H-ZISN, and $8$-H-ZISN are shown as examples. This indicates that tunable half-metallicity is typical for narrow H-ZISNs. The energy difference between the Fermi energy and spin up CBM is getting larger as $N$ increases. This observation implies that half-metallicity could be realized by a deeper $p$-type doping in wider H-ZISNs. As it is shown in Fig.~\ref{fig:label4} (d), the energy difference between the Fermi energy and CBM is essentially converged to around 250 meV for $N>10$, which corresponds to a $\sim$75\% filling of the spin down band. At the same time, the energy difference between CBM and spin-up VBM decreases almost linearly as width increases. It is expected that tunable half-metallicity disappears once CBM touches spin-up VBM in a wide enough H-ZISN. By extrapolation, one can find that the half-metal regime in H-ZISNs can only exist for widths $L \lesssim 11.3$~nm ($N\lesssim 32$). Experimentally, graphene nanoribbons (GRNs) with a width smaller than 10~nm have been fabricated by various experimental techniques \cite{ma2013recent}.

Up to now, edge magnetism of H-ZISN has been considered on the basis of HSE06 calculations. Although there is an indication of the ferromagnetism (or, strictly speaking, superparamagnetism taking into account 1D character of the system under consideration) in H-ZISNs, the magnetic ground state cannot be conclusively determined within DFT due to symmetry constrains imposed by periodic boundary conditions. More reliable description of magnetic properties can be approached using the Heisenberg picture, which assumes localization of magnetic moments. In order to use this approach, we first construct a TB model which describes the HSE06 electronic structure. We then map the TB Hamiltonian onto the Heisenberg model to determine exchange interactions \cite{rudenko2013exchange}. After that, we discuss magnetic properties of H-ZISNs in more details, including their temperature stability.

\section{tight-binding model}
\begin{figure}[ht]	
	\includegraphics[width=8.5cm]{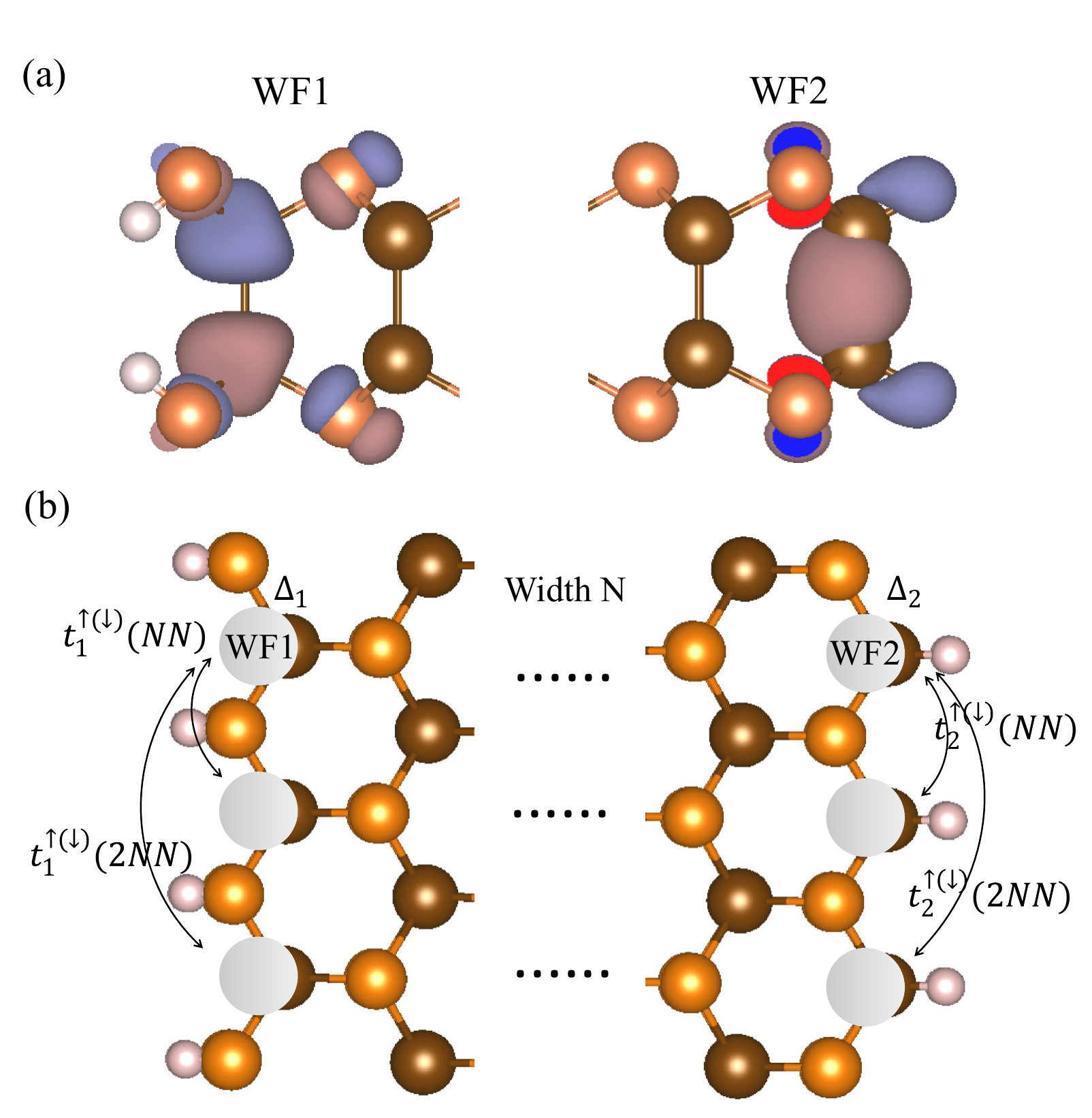}
	\caption{(a) Real-space distribution of WFs corresponding to the basis of the TB Hamiltonian for H-ZISNs, (b) Schematic representation of the most relevant hoppings $t_i^{\sigma}$ between the Wannier orbitals in the TB model for H-ZISN The orbitals are shown by gray labeled circles, localized at the corresponding WF centers. Each orbital is characterized by the spin splitting $\Delta_i=\varepsilon^{\uparrow}_i-\varepsilon^{\downarrow}_i$. Numerical parameters are listed in Table~\ref{table:label1}.}
	\label{fig:label5}
\end{figure}
We now turn to TB analysis of the band structure. Within DFT calculations, the magnetic moment stems mainly from the spin-splitting of the valence band (VB), while the conduction band (CB) remain spin degenerate. Here, we aim at a simplified description of the electronic structure of H-ZISNs taking only relevant electronic bands into account. To this end, we perform TB parametrization of the DFT Hamiltonian using one-dimensional (1D) two-band model,
\begin{equation}
H^{\sigma}=\sum_{i\sigma}\varepsilon_{i}^{\sigma}c^{\dagger}_{i\sigma}c_{i\sigma}+\sum_{ij\sigma }t^{\sigma}_{ij}c^{\dagger}_{i\sigma}c_{j\sigma},
\label{hamilt}
\end{equation}
where the summation runs over edge lattice sites, $\varepsilon_{i}^{\sigma}$ is the energy of an electron at site $i$ with spin $\sigma$=$\uparrow,\downarrow$, $c^{\dagger}_{i\sigma}~(c_{j\sigma})$ is the creation (annihilation) operator of electrons at site $i$ ($j$) with spin $\sigma$, and $t^{\sigma}_{ij}$ is the hopping parameter between sites $i$th and $j$th, corresponding to spin $\sigma$. Spin-orbit coupling is not included in the Hamiltonian Eq.~(\ref{hamilt}), meaning that there are no spin-flip processes ($c^{\dagger}_{i\uparrow}c_{j\downarrow}=0)$. Such terms have no direct effect on the isotropic exchange interaction (see Eq.~(\ref{exch}) below) and, therefore, are not relevant for our study.


\begin{table*}[ht]
	\caption{Parameters of the TB model for $N$-H-ZISN ($N$=5--12) as defined by Eq.~(\ref{hamilt}). $\Delta_i=\varepsilon_i^{\uparrow}-\varepsilon_i^{\downarrow}$ is the spin splitting of the orbital $i$ (in eV), and $t_{ij}^{\sigma}$ is the spin-dependent ($\sigma=\uparrow,\downarrow$) hopping parameter (in eV) between orbitals $i$ and $j$ as it is schematically shown in Fig.~\ref{fig:label5}. $\Omega^{\sigma}_i$ denotes quadratic spread (in~\AA$^{2}$) associated with the orbital $i$ and spin $\sigma$ (see text for the definition). NN and 2NN stand for the nearest-neighbor and second-nearest-neighbor interactions, respectively.
    }
	\begin{tabular}{cccccccccccc}
		\hline
		\multicolumn{1}{c}{Width $N$}&\multicolumn{4}{c}{WF1 (Se-edge)}& & \multicolumn{6}{c}{WF2 (In-edge)}\\
		\multicolumn{1}{c}{ }&\multicolumn{1}{c}{$\Omega_{1}^{\uparrow(\downarrow)}$}&\multicolumn{1}{c}{$\Delta_{1}$}&\multicolumn{1}{c}{$t^{\uparrow(\downarrow)}_{1}(\mathrm{NN})$}&\multicolumn{1}{c}{$t^{\uparrow(\downarrow)}_{1}(\mathrm{2NN})$} & &\multicolumn{1}{c}{$\Omega_{2}^{\uparrow}$}&\multicolumn{1}{c}{$\Omega_{2}^{\downarrow}$}& \multicolumn{1}{c}{$\Delta_{2}$}&\multicolumn{1}{c}{$t^{\uparrow}_{2}(\mathrm{NN})$}&\multicolumn{1}{c}{$t^{\downarrow}_{2}(\mathrm{NN})$}&\multicolumn{1}{c}{$t^{\uparrow(\downarrow)}_{2}(\mathrm{2NN})$}\\
		\hline
		5& 12.1 &0.00&	-0.14 &	-0.12 & &10.2&9.8&	0.23&	0.16 & 0.19&	-0.05 \\
		6&12.4&0.00&	-0.14 &	 -0.11& &10.9&10.1&0.29&	0.16 & 0.20&	-0.06 \\
		7&12.5&0.00&	-0.13 &	-0.11& &11.0&9.9&	0.36&	0.15 & 0.20&	-0.06 \\
		8&13.2&0.00&	-0.12 &	-0.11& &10.6&9.9&0.44&	0.16 & 0.21&	-0.06 \\
		9&13.3&0.00&	-0.12 &	-0.11& &11.8&10.0&0.44&	0.16 & 0.21&	-0.06 \\
		10&12.9&0.00&-0.12 & -0.11& &11.4&9.4&0.46&	0.16 & 0.22&	-0.06 \\
		12&12.8&0.00&	-0.12 &	-0.11& &11.1&9.1&0.47&	0.16 & 0.21&	-0.06 \\
		\hline
	\end{tabular}
    \label{table:label1}
\end{table*}

The parameters of the Hamiltonian defined above are determined by making use of the Wannier functions (WFs) \cite{marzari2012maximally}. Spin-polarized DFT electronic structure of a periodic material is characterized by the band dispersion $\epsilon^{\sigma}_{ik}$ and extended Bloch states $|\psi^{\sigma}_{nk}\rangle$, defined in terms of the band index $n$, crystal momentum $k$, and spin projection $\sigma$. Using the fact that there are two weakly interacting bands for each spin around the Fermi energy in H-ZISNs, we construct a set of two spin-dependent WFs $|w^{\sigma}_{i}(r)\rangle$ for $N$-H-ZISN ($N$=5--12). To this end, we do Fourier transform of the corresponding Bloch states, $|w^{\sigma}_{nR}\rangle=\sum_{k}e^{-ikR}|\psi_{nk}^{\sigma}\rangle$. The resulting WFs are schematically depicted in Fig.~\ref{fig:label5}(a). The Wannier orbitals (WF1 and WF2) are localized along the Se- and In-edge, respectively.
We then transform the DFT Hamiltonian in the WFs basis $\langle w^{\sigma}_{i}|H|w^{\sigma}_{j}\rangle$ and determine parameters appearing in Eq.~(\ref{hamilt}), i.e. the on-site energies $\varepsilon^{\sigma}_{i}$ and hoppings integrals $t^{\sigma}_{ij}$. 
For this purpose, we use the {\sc wannier90} code \cite{mostofi}.
The relevant parameters as well as WF spreads, $\Omega_i^{\sigma}=\langle w_{i}^{\sigma}|\hat{r}^{2}|w_{i}^{\sigma}\rangle-|\langle w_{i}^{\sigma}|\hat{r}|w_{i}^{\sigma} \rangle|^{2}$, are given in Table~\ref{table:label1}, while schematic representation of the TB model is shown in Fig.~\ref{fig:label5}(b). Already for $N=5$, the interaction between the H-ZISN edges is so weak that the corresponding hoppings can be neglected ($|t^{\sigma}_{ij}|<$0.01 eV). The hoppings beyond the second coordination sphere turn out to be negligible too. For $N=12$ all the parameters are almost converged. From Table~\ref{table:label1} one can also see that there is only one magnetic orbital (WF2) with nonzero spin-splitting $\Delta_2=\varepsilon_2^{\uparrow}-\varepsilon_2^{\downarrow}$, while the other orbital (WF1) is spin-degenerate ($\Delta_1=0$) independently of the ribbon width.

\begin{figure}[ht]	
	\includegraphics[width=8.5cm]{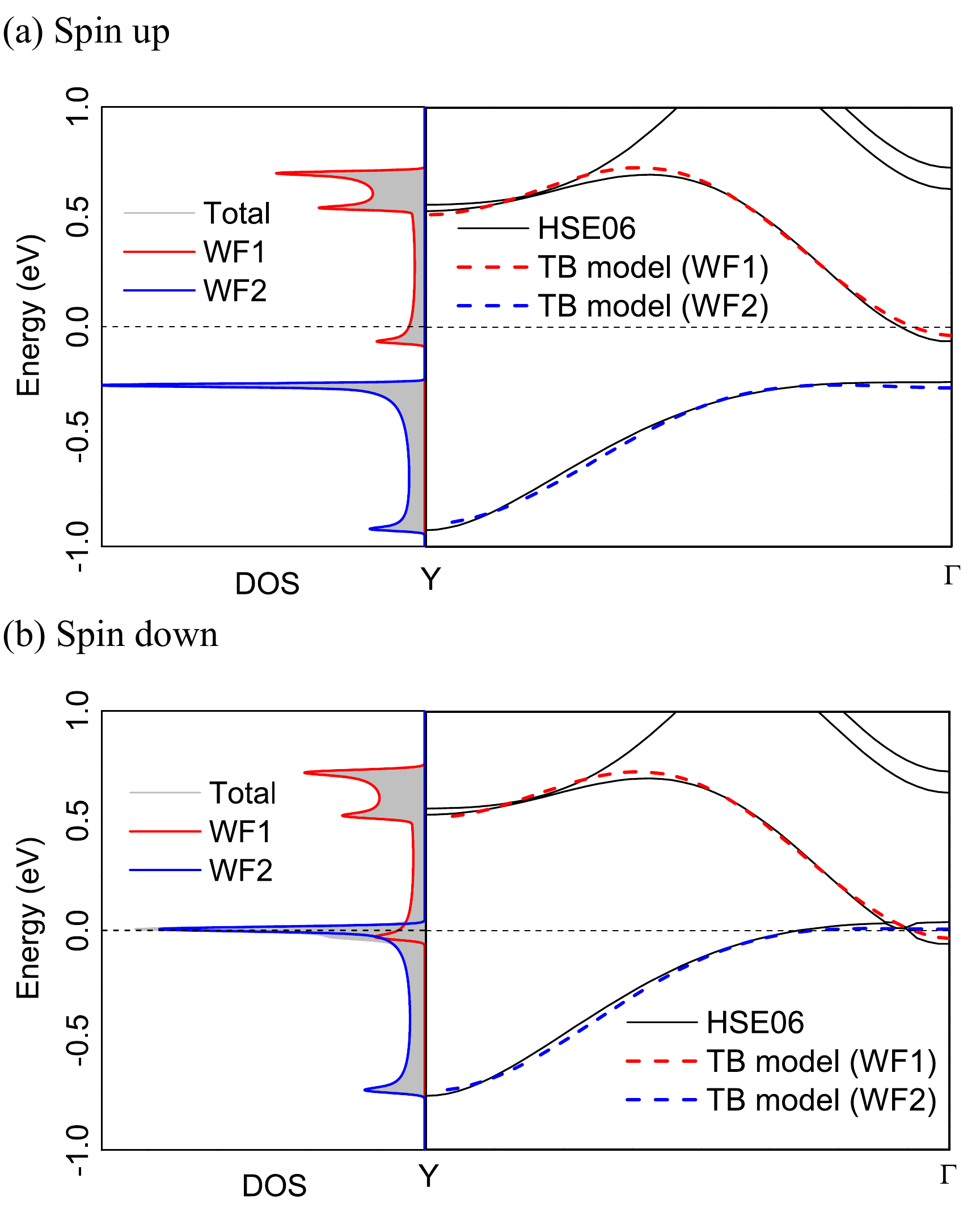}
	\caption{Comparison between the spin-resolved band structures of $5$-H-ZISN calculated using HSE06 (solid) and TB model (dash) [Eq.~(\ref{hamilt})]. Contribution from WFs is shown by different color: red for WF1 and blue for WF2 (see text for details). DOS is calculated from the TB model in the same way as described in the caption of Fig.~\ref{fig:label3}.}
	\label{fig:label6}
\end{figure}

As an example, band structure of 5-H-ZISN calculated from the two-band TB model is shown in Fig.~\ref{fig:label6}, which describes the HSE06 electronic structure near the Fermi level very well. Within the TB model, relevant bands can be projected on WFs, which is shown by color in Figs.~\ref{fig:label6}(a) and (b). One can see that WFs localized along the In-terminated edge contribute exclusively to the valence band, while WFs from the Se-terminated edge describe the conduction band. This is consistent with the PBE and HSE06 results presented earlier. The constructed TB model can be applied to estimate exchange interactions within the Heisenberg model, and for the calculation of transport properties.

\section{Heisenberg model}
Because of sufficiently strong localization of magnetic moments in H-ZISNs, the magnetic interactions can be analyzed in terms of the Heisenberg model,
\begin{equation}
	H=-\sum_{i\ne j}J_{ij}S_{i}S_{j},
\label{heisenberg}
\end{equation}
where $J_{ij}$ is the exchange coupling between spins at sites $i$ and $j$, and $S_{i(j)}$ is the unit vector pointing in the direction of the local magnetic moment at site $i$ ($j$). To calculate exchange interactions, we make use of the magnetic force theorem, which allows us to map the TB model introduced above onto the classical Heisenberg model. For a lattice with basis, exchange interactions can be written in the following form \cite{liechtenstein1987local},
\begin{equation}
	J^{\alpha\beta}_{ij}=\frac{1}{4\pi}\int_{-\infty}^{E_{f}}d\varepsilon \mathrm{Im}[\Delta_{\alpha}G^{\alpha \beta \downarrow}_{ij}(\varepsilon)\Delta_{\beta}G^{\beta \alpha \uparrow}_{ji}(\varepsilon)],
\label{exch}
\end{equation}
where $i,j$ and $\alpha , \beta $ are indices of the unit cell and orbitals, respectively. $\Delta_{\alpha}$ is the exchange splitting of $\alpha$th WF calculated from diagonal elements of the spin-polarized WF Hamiltonian as $\Delta_{\alpha}=H^{\uparrow}_{\alpha \alpha}-H^{\downarrow}_{\alpha \alpha}$, $E_{f}$ is the Fermi energy, and $G^{\alpha \beta \downarrow}_{ij}(\varepsilon)$ is the real-space Green's function, which can be obtained from its reciprocal counterpart via the Fourier transform,  $
	G^{\alpha\beta\sigma}_{ij}(\varepsilon)=\sum_{k}G^{\alpha\beta\sigma}_{k}(\varepsilon)e^{-i k (R_{i}-R_{j})}$.
In turn, the reciprocal-space Green's function reads (in the matrix form) 
\begin{equation}
	G^{\sigma}_{k}(\varepsilon)=[\varepsilon-H^{\sigma}(k)+i\eta]^{-1},
\end{equation}
where $H^{\sigma}(k)$ is the reciprocal Hamiltonian, whose matrix elements can be easily obtained from the TB model, Eq.~(\ref{hamilt}). To calculate the reciprocal-space Green's function, we use $10^{4}$ k-points and $\eta=0.1$~meV.

\begin{table}
	\caption{Exchange interactions calculated up to the third-nearest-neighbor (3NN) for undoped $N$-H-ZISN ($N$=5--12) by using the magnetic force theorem [Eq.~\ref{exch}]. The parameters listed are numerically accurate to within 0.01 meV.}
	\begin{tabular}{cccc}
		\hline
		Width $N$&	NN~(meV)&	2NN~(meV)&	3NN~(meV)\\
		\hline
		5&	2.12&	$-$0.15&	$-$0.08\\
		6&	2.91&	$-$0.22&	0.05\\
		7&	3.53&	$-$0.26&	0.10\\
		8&	4.45&	$-$0.16&	0.21\\
		9&	4.27&	$-$0.16&	0.19\\
		10&	4.56&	$-$0.11&	0.23\\
		12&	4.76&	0.15&	0.24\\
		\hline
	\end{tabular}
    \label{table:label2}
\end{table}

\begin{figure}[!htp]	
	\includegraphics[width=8.5cm]{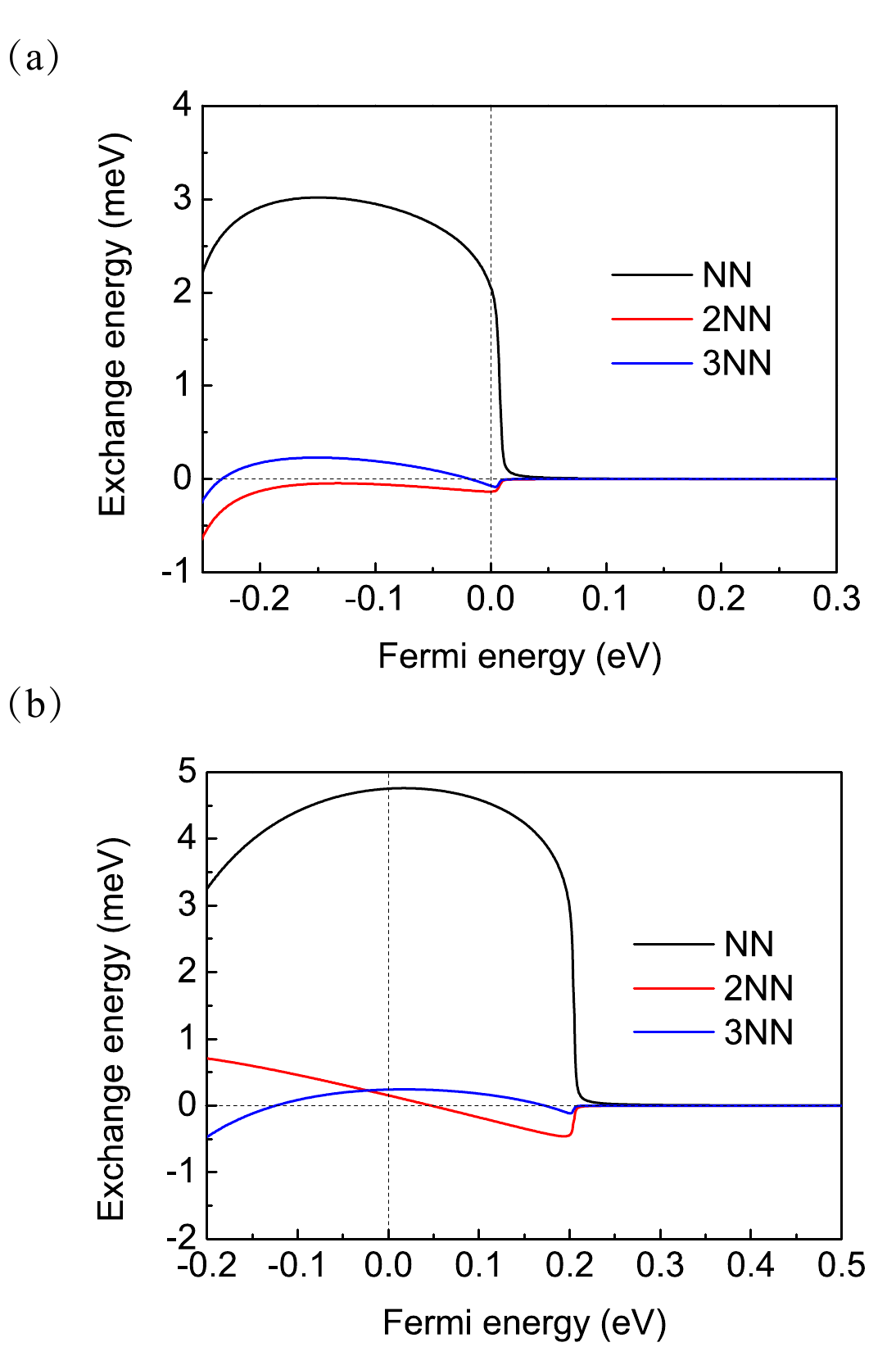}
	\caption{Exchange interactions for $5$-H-ZISN (a) and $12$-H-ZISN (b) calculated as a function of the Fermi energy shown up to the third coordination sphere.}
	\label{fig:label7}
\end{figure}

The magnetic lattice of H-ZISNs can be simplified to an 1D magnetic chain because magnetic moments entirely localized along one of the edges. The resulting exchange interactions calculated using Eq.~(\ref{exch}) up to the third coordination sphere are summarized in Table~\ref{table:label2} for different ribbon widths. One can see that the leading exchange interaction $J_\mathrm{NN}$ is FM, which depends considerably on the ribbon width $N$, ranging from $J_{\mathrm{NN}}\approx 2$ meV ($N$=5) to $J_{\mathrm{NN}}\approx 5$ meV ($N=12$). Similar to the magnetic moments, $J_{\mathrm{NN}}$ is almost converged with the ribbon width at $N=12$. The more distant interactions (2NN and 3NN) are smaller by more than an order of magnitude. One can see, however, that $J_\mathrm{2NN}$ and $J_\mathrm{3NN}$ are comparable in magnitude, whereas their signs are different for most $N$s. This behavior is typical to the RKKY interactions, which is not unexpected taking into account that H-ZISNs are conductors at zero doping. In Figs.~\ref{fig:label7} (a) and (b), we show how the exchange interactions depend on the Fermi energy considering the examples of 5-H-ZISN and 12-H-ZISN. In the first case [$N=5$, Fig.~\ref{fig:label7}(a)], a rapid decay of the exchange interaction can be observed for positive Fermi energies ($n$-doping). On the contrary, the exchange interactions monotonously increase for negative Fermi energies ($p$-doping) until $E_f\approx -0.15$ eV. In the second case [$N=12$, Fig.~\ref{fig:label7}(b)], $J_{\mathrm{NN}}$ has its maximum at zero Fermi energy and exchange interactions decrease rapidly down to zero at positive Fermi energy around $0.2$ eV. In this case, both $n$- and $p$-doping slightly reduce $J_{\mathrm{NN}}$. From Fig.~\ref{fig:label7} one can also see how $J_{\mathrm{2NN}}$ and $J_{\mathrm{3NN}}$ change their sign as function of the Fermi energy. Overall, in the context of doping-induced half-metallicity in H-ZISNs discussed earlier, exchange interactions are not significantly affected by $p$-type doping.

\begin{figure}[ht]	
	\includegraphics[width=8.5cm]{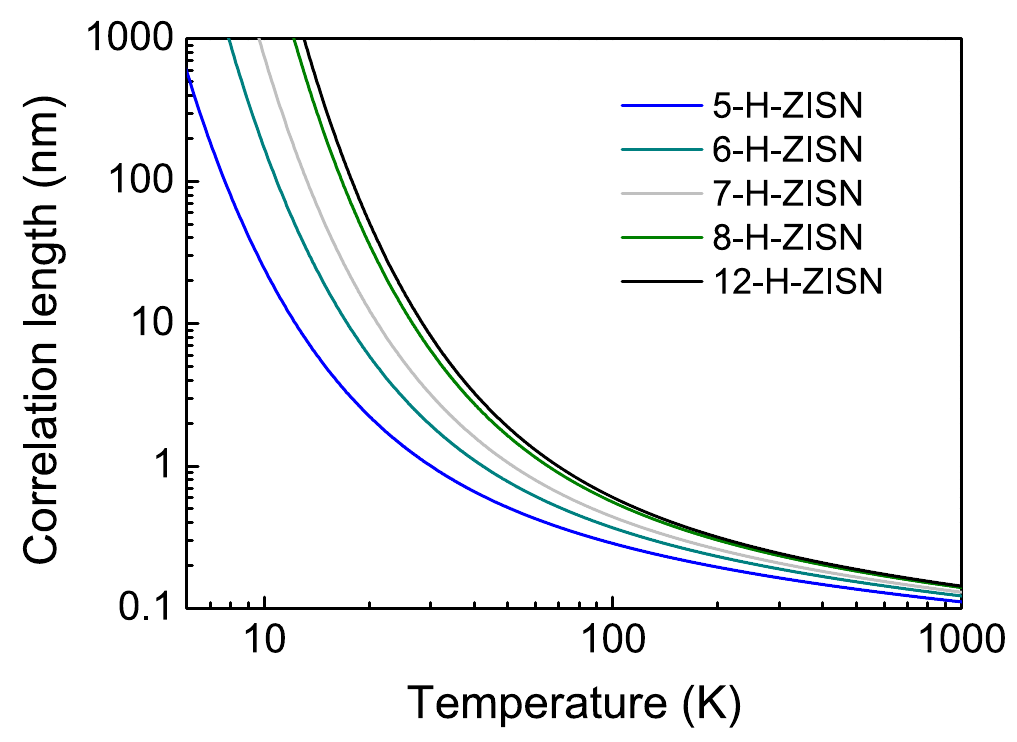}
	\caption{Spin correlation length of $N$-H-ZISN calculated within the Ising model for different widths ($N$=5--12) as a function of temperature.}
	\label{fig:label8}
\end{figure}

Having obtained the exchange interactions between magnetic moments, we can determine magnetic ground state of H-ZISNs. Compared to the dominant FM interaction $J_{\mathrm{NN}}$, other interactions are essentially negligible. In this situation, one can expect either FM ground state or spin spiral state with a large spiral vector. In order to check the stability of FM magnetic configuration, we take 2NN interaction into account, which is AFM for narrow H-ZISN. To this end, we consider the total energy of the coplanar spin spiral $H(\theta)=\sum_{ij}J_{ij}{\bf S}_i(\phi_i){\bf S}_j(\phi_i+\theta)$ as function of the phase shift between spins, $\theta$. We then minimize $H(\theta)$ with respect to $\theta$, and find that the ground state corresponds to $\theta=0^{\circ}$, which corresponds to the FM case. The result is not surprising because for one-dimensional Heisenberg chain with $J_{1}-J_{2}$, where $J_{1}$ is FM interaction and $J_{2}$ is AFM interaction, the ground state is FM when $|\frac{J_{2}}{J_{1}}|<0.25$ \cite{hamada1988exact}. Therefore, disregarding temperature effects and long-wavelength fluctuations, the ground state magnetic ordering of H-ZISN is FM.

\section{Correlation length}
Up to now magnetism in H-ZISNs has been discussed without considering temperature, which induces fluctuations of magnetic moments. Temperature effects are especially important for magnetic nanoribbons since there is no long-range magnetic order in 1D at any finite temperature \cite{mermin1966absence}. Generally, the range of magnetic order is characterized by the correlation length $\xi$, which defines the decay law of the spin correlation function with distance, $\langle S_iS_j \rangle = \langle S_iS_i \rangle \mathrm{exp} (-|r_i - r_j|/\xi)$ \cite{yazyev2008magnetic}. To obtain the correlation length, we start from the 1D Heisenberg model given by Eq.~(\ref{heisenberg}). Taking into account the dominant role of NN exchange interaction ($J_{\mathrm{NN}}\gg J_{\mathrm{2NN}}$) and neglecting intersite magnetic anisotropy ($\langle S_iS_i \rangle=1$), the spin Hamiltonian can be simplified to the form of 1D Ising model, $H=-J_{\mathrm{NN}}\sum_{i}S^z_iS^z_{i+1}$. In the thermodynamic limit, the corresponding correlation length in zero field can be written as \cite{stanley1971phase}
\begin{equation}
	\xi=\left[\mathrm{ln}\left(\mathrm{tanh}\left(\frac{J_{\mathrm{NN}}}{k_{B}T}\right)\right)\right]^{-1}.
\end{equation}
We evaluate the zero-field spin correlation length in H-ZISN as a function of temperature and the ribbon width $N$. As can be seen from Fig.~\ref{fig:label8}, the correlation length increases for larger $N$ and converges to a constant value at $N\approx 12$. The different correlation length observed for varying ribbon width is related to the difference in exchange interactions (see Table \ref{table:label2}). At room temperature ($300$~K), the largest ($N$=12) correlation length $\xi=0.6$ unit cell ($\approx 0.3$~nm), meaning that spintronics devices based on magnetic H-ZISNs can not be operated at room temperature if its length is beyond the atomic scale. At liquid nitrogen temperature ($\approx 77$~K), the correlation length is considerably larger, $\xi=2.3$ unit cells ($ \approx 1$~nm). The size of a device could be extended beyond the micrometer scale at low enough temperature, e.g. liquid helium temperature ($ \approx 4.4$~K). Although the zero-field correlation lengths are not particularly large, the magnetic stability of the nanoribbons can be further enhanced by the application of external magnetic field, or by a substrate with strong spin-orbit coupling.

\section{Spin-dependent Transport}
To gain insight into the role of magnetism in the transport properties of H-ZISNs, we calculate the spin-dependent electronic conductivity as function of chemical potential (doping). The expression for dc conductivity as function of chemical potential $\mu$ and temperature $T$ reads \cite{scheidemantel2003transport}:

\begin{equation}
	\sigma(\mu,T)=\frac{e^{2}}{V}\sum_{n,k}\int dE \left(-\frac{\partial f(E,\mu,T)}{\partial E}\right)v^2_{nk}\tau_{nk}\delta(E-E_{nk})
\end{equation}
where the summation runs over all relevant bands $n$ and $k$-points of the Brillouin zone. $V=L_xL_yL_z$ is effective nanoribbon volume with $L_x$ and $L_y$ being the ribbon width and lattice constant along the periodic direction, and $L_z$=8.32~\AA~\cite{olguin2013ab} is the effective thickness of a single-layer in InSe bulk crystal. $E_{nk}$ and $v_{nk}$ is the corresponding band energy and group velocity, and $f(E,\mu,T)=(\mathrm{exp}[(E-\mu)/T]+1)^{-1}$ is the Fermi-Dirac distribution function, where we take $T=300$~K in all calculations. In our case, group velocity is nonzero only along one (periodic) direction. $\tau_{nk}$ is the carrier lifetime, for which we adopt the relaxation time approximation and assume that the lifetime $\tau_{nk}$ is independent of both $n$ and $k$, i.e. $\tau_{nk}=\tau$. Here, we do not specify scattering mechanism, which may include scattering on defects \cite{yuan2015transport} or phonons \cite{rudenko2016intrinsic}. To assume a reasonable lifetime, we estimate the group velocity $v_{nk}=\frac{1}{\hbar}\frac{\partial E_n(k)}{\partial k}$ from the TB band structure. The resulting group velocities are found to be on the order of $10^{5}-10^{6}$~m/s. Using the approximate relation for the mean free path $\xi_l \approx v \cdot \tau$, we choose $\tau=10$~fs to ensure $\xi_l \sim 1$~nm. In order to calculate $\sigma(\mu,T)$ for different ribbon widths, we use the {\sc boltzwann} code \cite{pizzi2014boltzwann} in conjunction with WFs obtained previously (see Sec.~III). 

\begin{figure}[!pht]	
	\includegraphics[width=8.5cm]{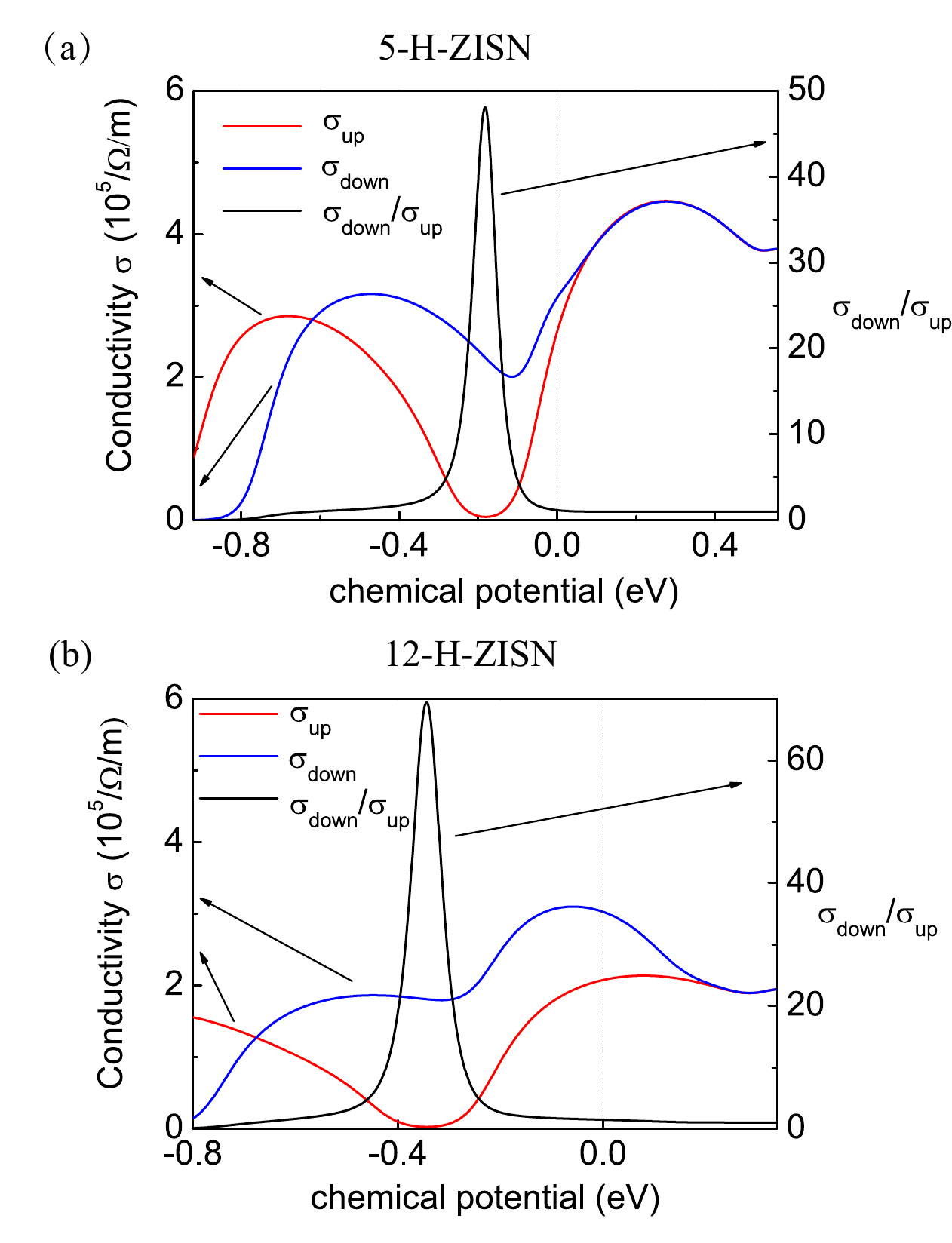}
      \caption{Electrical conductivity calculated for 5-H-ZISN (a) and 12-H-ZISN (b) for spin up (red lines) and spin down (blue lines) conduction channels, as well as the ratio between them (black lines) shown as function of chemical potential. The calculations are performed at $T=300$~K.}
	\label{fig:label9}
\end{figure}

Spin-dependent conductivity calculated for two representative cases, 5-H-ZISN and 12-H-ZISN, is shown in Fig.~\ref{fig:label9}. Overall, the conductivities resemble DOS as there are no anomalies in the group velocity. At zero chemical potential, both spin-up and spin-down electrons have comparable contribution to the conductivity independently of the ribbon width. At positive chemical potential ($n$-doping) magnetism in H-ZISNs disappears (see Fig.~\ref{fig:label7}) and $\sigma_{\mathrm{down}}/\sigma_{\mathrm{up}}\rightarrow 1$. The situation for negative chemical potential ($p$-doping) is different. Due to the valence band splitting, there is an energy region in which $\sigma_{\mathrm{up}}\approx 0$, resulting in a sharp peak of $\sigma_{\mathrm{down}}/\sigma_{\mathrm{up}}$. Therefore, in this regime the system behaves as a perfect spin filter. The corresponding (critical) chemical potentials depend on the ribbon width, ranging from $-$0.4 eV ($N$=12) to $-$0.2 eV ($N$=5). The dependence of the electronic conductivity on the spin channel in H-ZISNs makes these system promising candidates for the realization of spintronic effects at the nanometer scale.

\section{Conclusion}
We have systematically studied the electronic and magnetic properties of ZISNs with edges saturated by hydrogen. By performing hybrid-functional first-principles calculations, we find that H-ZISNs are materials with tunable half-metallicity and short-range magnetic order. Properties of H-ZISNs are demonstrated to be highly susceptible to charge doping. Particularly, $p$-type doping turns semimetallic H-ZISNs into half-metal with spin-polarization emerging along In-terminated edge. On the contrary, $n$-type doping results in a spin-degenerate (nonmagnetic) ground state. To analyze magnetic interactions in H-ZISNs, we construct a tractable TB model and perform a mapping to the Heisenberg model. We find that the dominant interaction in H-ZISNs is ferromagnetic, whose magnitude considerably increases with the ribbon width. This behavior appears interesting for spintronic applications, for instance, in the context of spin filtering.
Indeed, our analysis of spin-dependent electronic transport reveals the existence of a regime with spin-selective dc conductivities. The feasibility of such applications is primarily determined by the spin correlation length. Our estimation leads to 0.3 nm for room temperatures, and to 1 nm for liquid nitrogen temperatures. Further enhancement of magnetic stability could be achieved by the presence of a substrate, or external magnetic field. Our findings is a step forward toward the understanding of magnetism in low-dimensional materials, which can stimulate further theoretical and experimental studies.

\begin{acknowledgments}
This work is supported by the National Key R$\&$D Program of China (Grant No.2018FYA0305800) and National Science Foundation of China (Grant No. 11774269). The authors would like to thank M. I. Katsnelson for helpful discussions. Numerical calculations presented in this paper have been performed on a supercomputing system in the Supercomputing Center of Wuhan University.
\end{acknowledgments}

\bibliographystyle{apsrev4-1}
\bibliography{references}

\end{document}